\documentclass[12pt]{article}
\usepackage{amssymb}

\input{tcilatex}

\begin{document}

\begin{center}
\textbf{THOUGHTS ON DUALITY AND}

\smallskip\ 

\textbf{FUNDAMENTAL CONSTANTS}

\textbf{\ }

J. A. Nieto \footnote{%
nieto@uas.uasnet.mx}, L. Ruiz and J. Silvas

\smallskip

\textit{Facultad de Ciencias F\'{\i}sico-Matem\'{a}ticas, Universidad Aut%
\'{o}noma}

\textit{de Sinaloa, C.P. 80000, Culiac\'{a}n, Sinaloa, M\'{e}xico}

\smallskip\ 

\textbf{Abstract}
\end{center}

{\small We consider some fundamental constants from the point of view of the
duality symmetry. Our analysis of duality is focused on three issues: the
maximum radiated power of gravitational waves, the cosmological constant,
and the magnetic monopole mass. We show that the maximum radiated power of
gravitational waves implies that the Planck time is a minimal time.
Furthermore, we prove that duality implies a quantization of the
cosmological constant. Finally, by using one of the Euler series for the
number }$\pi $,{\small \ we show that the Dirac electric-magnetic charge
quantization implies a mass for the magnetic monopole (or neutrino) of the
order of }$10^{-5}${\small \ the mass of the electron.}

\begin{center}
\smallskip\ \textbf{Resumen}
\end{center}

{\small Consideramos algunas constantes fundamentales desde el punto de
vista de la simetr\'{\i}a de dualidad. Nuestro an\'{a}lisis de dualidad se
enfoca en tres temas: la potencia m\'{a}xima radiada de ondas
gravitacionales, la constante cosmol\'{o}gica y la masa del monopolo magn%
\'{e}tico. Demostramos que la potencia m\'{a}xima radiada de ondas
gravitacionales implica que el tiempo de Planck corresponde a un tiempo m%
\'{\i}nimo. M\'{a}s aun, probamos que la dualidad implica una cuantizaci\'{o}%
n de la constante cosmol\'{o}gica. Finalmente, usando una de las series de
Euler para el n\'{u}mero }$\pi ${\small , demostramos que la cuantizaci\'{o}%
n de la carga el\'{e}ctrica-magn\'{e}tica de Dirac implica una masa para el
monopolo magn\'{e}tico (o neutrino) del orden de }$10^{-5}${\small \ la masa
del electr\'{o}n.}

{\small \ }


Pacs numbers: 04.60.-m, 04.30.-w, 6.20.Jr, 98.80.-k

September, 2006

\newpage \noindent \textbf{1.- Introduction}

\smallskip\ 

Because the problem of the number of fundamental constants [1] and their
possible time variability [2] is of permanent interest in physics, any
consistent new idea on this subject must be welcome. In this context, it has
been emphasized [3] that one should only consider as physically meaningful
the variability of dimensionless constants rather than dimensional constants
[2]. This claim is not shared, however, by some physicists (see Ref. 2 for
details), and therefore new routes for approaching the subject seem to be
needed.

One of our aims in this paper is to shed some light on the above controversy
by applying the duality concept to some fundamental constants. Specifically,
in this work, we analyze some fundamental constants from the point of view
of a duality symmetry, including the Planck time, the cosmological constant,
and the magnetic monopole mass. We show that by applying the duality concept
to the maximum radiated power of gravitational waves one obtains the result
that the Planck time must be a minimal time. Furthermore, using the $S$%
-duality concept for the cosmological constant, obtained in the linearized
gravity\ development [4], and relaying on analogy of the Dirac's
quantization of the electric and magnetic monopole charges, we argue that
duality implies a quantization of the cosmological constant. Finally, by
using one of the Euler series for the number $\pi $, we demonstrate that the
Dirac duality concept for the electric charge implies a relation between the
electron mass $m_{e}$ and the magnetic monopole mass $m_{g}$. Such a
relation leads to a value for $m_{g}$ of the order of the neutrino mass $%
\sim 10^{-5}$ $m_{e}$, which is too low in comparison with the expected
standard value for the mass of the magnetic monopole, namely of the order of 
$GeVs$. Thus, we find that duality seems to imply a deep connection between
the neutrino $\bar{\nu}_{e}$ and the magnetic monopole.

Moreover, we explain that the three different types of results mentioned
above can be written in a dimensionless constant context. This suggests that
the underlying theory must be invariant under the duality of the
dimensionless fundamental constants rather than a duality of dimensional
constants. This result is in agreement with Dirac's older idea [3] (see Ref.
2 for a recent discussion of this problem) that dimensionless constants are
more important than dimensional ones. From this perspective, one may
conclude from our results that in fact what matters is the variability of
dimensionless fundamental constants, as Duff has emphasized [2], rather than
the variability of dimensional fundamental constants.

This article is organized as follows. In Sec. 2, using the maximum radiated
power of gravitational waves, we prove that the Planck time is a minimal
time. In Sec. 3, we discuss the cosmological constant duality, and in Sec. 4
we analyze the magnetic monopole mass from a duality perspective. Finally,
in Sec. 5, we make some latter remarks.

\bigskip\ 

\noindent \textbf{2.- Duality between the maximum radiated power and Planck
time}

\smallskip\ 

Consider a source of gravitational waves of mass $M$ and radius $r$. It is
known that an estimate of the radiated power of gravitational waves is given
by%
\begin{equation}
P\sim L_{0}(\frac{r_{Sch}}{r})^{5},  \label{1}
\end{equation}%
where

\begin{equation}
L_{0}=\frac{c^{5}}{G},  \label{2}
\end{equation}%
and

\begin{equation}
r_{Sch}=\frac{GM}{c^{2}}.  \label{3}
\end{equation}%
Here, $c$ is the "light" velocity (or spacetime structure constant in the
terminology of Ref. 5) and $G$ is the Newton gravitational constant. In
order to avoid the collapse of the object into a black hole, it is necessary
to have $r_{Sch}<r$ and therefore from formula (1) we see that the maximum
radiated power of any object is $L_{0}$. Conversely, if we assume that $%
L_{0} $ is the maximum radiated power, then from (1) we obtain the relation $%
r_{Sch}<R$, which is linked to the relation $v<c$, where $v$ is the velocity
of the source.

Let us now introduce the Planck time

\begin{equation}
t_{P}=(\frac{G\hslash }{c^{5}})^{1/2},  \label{4}
\end{equation}%
where $\hslash $ is the Planck constant. This formula can be written as

\begin{equation}
\frac{\hslash }{t_{P}^{2}}=\frac{c^{5}}{G}=L_{0}.  \label{5}
\end{equation}%
Therefore, by fixing $\hslash $, we obtain the interesting dual property: $%
L_{0}$ is the maximum radiated power if and only if $t_{P}$ is a minimal
time. Of course, when $c$ is setting, one has that minimal time $t_{P}$
implies that the Planck length $l_{P}=ct_{P}=(\frac{G\hslash }{c^{3}})^{1/2}$
is a minimum length in nature (see Ref. 6). Although, this result seems to
be in agreement with the idea that a fundamental length arose in the string
theory (see Ref. 7), its classical derivation presented here contrasts with
the same result obtained from quantum gravity (see Refs. 8 to 11, and
references. therein).

\bigskip\ 

\noindent \textbf{3.- Cosmological constant duality }

\smallskip\ 

In Ref. 4 it was proved that linearized gravity \textit{a la}
MacDowell-Mansouri implies a cosmological constant duality symmetry

\begin{equation}
\Lambda \leftrightarrow \frac{1}{\Lambda },  \label{6}
\end{equation}%
which can be thought as the analogue of the charge duality in an Abelian
gauge field theory,

\begin{equation}
e^{2}\leftrightarrow \frac{1}{e^{2}}.  \label{7}
\end{equation}%
In order to clarify this analogy, let us briefly describe the main result of
Ref. 4. Let us introduce the `gauge' field of linearized gravity,%
\begin{equation}
A_{\mu \alpha \beta }=\frac{1}{2}(\partial _{\alpha }h_{\mu \beta }-\partial
_{\beta }h_{\mu \alpha })=-A_{\mu \beta \alpha }.  \label{8}
\end{equation}%
Under the transformation%
\begin{equation}
\delta A_{\mu \alpha \beta }=\partial _{\mu }\lambda _{\alpha \beta },
\label{9}
\end{equation}%
the curvature tensor%
\begin{equation}
F_{\mu \nu }^{\alpha \beta }=\partial _{\mu }A_{\nu }^{\alpha \beta
}-\partial _{\nu }A_{\mu }^{\alpha \beta }  \label{10}
\end{equation}%
is invariant. This means that the tensor $F_{\mu \nu }^{\alpha \beta }$ can
be identified with an abelian field strength.

Consider the extended curvature%
\begin{equation}
\mathcal{F}_{\mu \nu }^{\alpha \beta }=F_{\mu \nu }^{\alpha \beta }+\Omega
_{\mu \nu }^{\alpha \beta },  \label{11}
\end{equation}%
where

\begin{equation}
\Omega _{\mu \nu }^{\alpha \beta }=\delta _{\mu }^{\alpha }h_{\nu }^{\beta
}-\delta _{\mu }^{\beta }h_{\nu }^{\alpha }-\delta _{\nu }^{\alpha }h_{\mu
}^{\beta }+\delta _{\nu }^{\beta }h_{\mu }^{\alpha }.  \label{12}
\end{equation}%
In Ref. 4 it was shown that the action

\begin{equation}
\mathcal{S}=\frac{1}{16\Lambda }\int d^{4}x{}\epsilon ^{\mu \nu \alpha \beta
}\mathcal{F}_{\mu \nu }^{\tau \lambda }\mathcal{F}_{\alpha \beta }^{\sigma
\rho }{}\epsilon _{\tau \lambda \sigma \rho }+\frac{i\Theta }{8\pi }\int
d^{4}x{}\epsilon ^{\mu \nu \alpha \beta }{}\mathcal{F}_{\mu \nu }^{\tau
\lambda }{}\mathcal{F}_{\alpha \beta }^{\sigma \rho }{}\delta _{\tau \lambda
\sigma \rho },  \label{13}
\end{equation}%
where $\Lambda $ and $\Theta $ are constants, permits a dual action. From
(13) we observe that the cosmological constant $\Lambda $ is playing the
role of a gauge coupling constant $g^{2}$, and that $\Theta $ is playing the
role of a $\theta $ constant in the usual abelian Maxwell theory. Thus, we
find that the analogue of the gauge coupling constant duality $%
g^{2}\rightarrow \frac{1}{g^{2}}$ in the case of linearized gravity
corresponds to the cosmological constant duality transformation $\Lambda
\rightarrow \frac{1}{\Lambda }$ (see Ref. 4 for details).

In this section we are interested in a deep understanding of the relation
(6). For this purpose let us recall how the relation (7) arises in an
Abelian gauge field theory. It turns out that the origin of (7) is the
Dirac's electric charge quantization condition, namely

\begin{equation}
ge=\frac{n\hslash c}{2},  \label{14}
\end{equation}%
where $g$ is the magnetic monopole charge. The key point is that the
source-free Maxwell field equations are invariant under the transformation%
\begin{equation}
\begin{array}{c}
E\rightarrow B \\ 
\\ 
B\rightarrow -E.%
\end{array}
\label{15}
\end{equation}%
While in the case of nonsource-free Maxwell equations the transformation (9)
needs to be extended and accompanied by the transformation%
\begin{equation}
g\leftrightarrow e.  \label{16}
\end{equation}%
Due to (14), one sees that (16) is equivalent to (7).

In general, the cosmological constant $\Lambda $ can be written in terms of
a fundamental length $l$ in the form

\begin{equation}
\Lambda =\pm \frac{(D-1)(D-2)}{2l^{2}},  \label{17}
\end{equation}%
where $D$ is the dimension of the spacetime of an arbitrary signature.
Therefore, the duality relation (6) is equivalent to%
\begin{equation}
l^{2}\leftrightarrow \frac{1}{l^{2}}.  \label{18}
\end{equation}%
We observe that (18) establishes the analogy between (6) and (7) in a
clearer context. Thus, following this analogy, one should expect (18) to be
a consequence of the quantization relation%
\begin{equation}
\mathcal{L}l=\frac{nl_{p}R}{2},  \label{19}
\end{equation}%
where $l_{p}$ is the Planck length, $R$ is the radius of the universe and $%
\mathcal{L}$ is the dual length associated with $l$. In turn, this result
implies a quantization of $l$, and therefore a quantization of the
cosmological constant via the relation (17). In fact, by writing $\Lambda
_{l}\equiv \Lambda $ and%
\begin{equation}
\Lambda _{\mathcal{L}}=\pm \frac{(D-1)(D-2)}{2\mathcal{L}^{2}},  \label{20}
\end{equation}%
we discover that (19) implies the formula

\begin{equation}
\Lambda _{\mathcal{L}}\Lambda _{l}=\frac{(D-1)^{2}(D-2)^{2}}{%
n^{2}l_{p}^{2}R^{2}}.  \label{21}
\end{equation}%
Of course, the cases $D=1$ and $D=2$ are exceptional, as can be seen even
from (17). So, out of these two cases, one may be interested in an
understanding of the meaning of (19) and (21). First of all, if $\Lambda _{%
\mathcal{L}}\neq 0$, we discover that $\Lambda _{l}$ should be quantized.
Second, assuming $\mathcal{L\sim }\frac{R}{2}$, we observe from (19) that $%
l=nl_{p}$ and therefore $l_{p}$ is a minimal length, in agreement with our
discussion in Sec. 3. Finally, from (19) we see that, taking $\mathcal{L\sim 
}\frac{l_{p}}{2}$, one obtains $l=nR$, and therefore from (17) or (21) we
find that

\begin{equation}
\Lambda _{l}=\pm \frac{(D-1)(D-2)}{2n^{2}R^{2}}.  \label{22}
\end{equation}%
For $n=1,D=4$ and $R\sim 10^{28}cm$ we get $\Lambda _{l}\sim 10^{-5\text{ }%
6}cm^{-2}$, which is a very small value but nevertheless different from
zero. It is not difficult to see that these results can be dualized, that
is, when $\Lambda _{l}$ is small, $\Lambda _{\mathcal{L}}$ is large and vice
versa. For historical reasons the attempt to make zero the cosmological
constant is called "the cosmological constant problem". From (21) we observe
that for $D\neq 1$ and $D\neq 2$, this type of problem has no a solution
free of singularities. In fact, (21) implies that if $\Lambda
_{l}\rightarrow 0$, then $\Lambda _{\mathcal{L}}\rightarrow \infty $\ and
vice versa.

\bigskip\ 

\noindent \textbf{4.- The magnetic monopole mass duality}

\smallskip\ 

Consider the duality transformations 
\begin{equation}
g^{2}\longleftrightarrow \frac{1}{e^{2}}  \label{23}
\end{equation}%
and

\begin{equation}
m_{g}\longleftrightarrow \frac{1}{m_{e}}.  \label{24}
\end{equation}%
Observe that (23) is a consequence of (14). In (24), $m_{g}$ refers to the
mass of the magnetic monopole. Moreover, we are assuming that there exist
the analogue of the formula (24) for the mass quantization as Zee [12] has
suggested for any massive system. It is not difficult to see that the
relation

\begin{equation}
\frac{m_{g}g^{2}}{m_{e}e^{2}}=\beta ,  \label{25}
\end{equation}%
is invariant under the transformations (23) and (24). Thus, the constant $%
\beta $ in (25) must be fundamental, dimensionless, and should not be
related to any property of the system. On the other hand, it is known that
not only the fine structure constant $\alpha =\frac{e^{2}}{\hslash c}$ can
be related to the number $\pi $ via the Weyler heuristic formula

\begin{equation}
\alpha =\frac{9}{8\pi ^{4}}\left( \frac{\pi ^{5}}{2^{4}5!}\right) ^{1/4},
\label{26}
\end{equation}%
but also all masses of fundamental particles via the hiperdiamons lattices
based on Clifford algebras (see Ref. 13 and references therein). This
suggests that $\beta $ in (25) could, in principle, be related to the number 
$\pi .$ Let us choose one of the simplest possibilities for such a constant,
namely $\beta =a\pi ^{2}$, where $a$ is a numerical factor independent of $%
\pi $ to be determined below. Thus expression (25) becomes

\begin{equation}
\frac{m_{g}g^{2}}{m_{e}e^{2}}=a\pi ^{2}.  \label{27}
\end{equation}%
Using (14) and the fine structure constant $\alpha =\frac{e^{2}}{\hslash c}$%
, formula (27) yields

\begin{equation}
m_{g}=4am_{e}\alpha ^{2}\pi ^{2}.  \label{28}
\end{equation}%
It turns out to be convenient to multiply this expression by $c^{2}$

\begin{equation}
m_{g}c^{2}=4am_{e}c^{2}\alpha ^{2}\pi ^{2}.  \label{29}
\end{equation}

On the other hand, there exists a famous numerical series due to Euler for
determining the number $\pi $, namely

\begin{equation}
\sum_{n=1}^{\infty }\frac{1}{n^{2}}=\frac{\pi ^{2}}{6}.  \label{30}
\end{equation}%
which can be used in the Eq. (29) to obtain the intriguing result

\begin{equation}
m_{g}c^{2}=\sum_{n=1}^{\infty }\frac{m_{e}c^{2}\alpha ^{2}}{2}\frac{1}{n^{2}}%
,  \label{31}
\end{equation}%
provided we set $a=\frac{1}{2(4!)}.$ Therefore, we have shown that using
(14) the invariant formula (25) with $\beta =\frac{\pi ^{2}}{2(4!)}$ leads
to (31). We recognize in the expression%
\begin{equation}
E_{n}\equiv -\frac{m_{e}c^{2}\alpha ^{2}}{2}\frac{1}{n^{2}}  \label{32}
\end{equation}%
the well known formula for the eingenvalues of the energy for the hydrogen
atom. From (31) we find that the value of $m_{g}$ is of the order of the
neutrino mass, $m_{\nu _{e}}\sim 10^{-5}m_{e},$ but too low in comparison
with the expected standard value for the magnetic monopole mass, which is of
the order of $GeVs$. One may try to understand this result by considering
the well known neutron decay

\begin{equation}
n\rightarrow p+e+\bar{\nu}_{e}.  \label{33}
\end{equation}%
A hydrogen atom is made out of a proton $p$ and an electron $e$. Thus, the
transition (33) suggests that the total energy obtained by the eigenvalues
of the energy according to (32) should determine the mass of the neutrino $%
\bar{\nu}_{e}$. However, relation (32) suggests identifying $m_{\upsilon }$
with $m_{g},$ and therefore, we may conclude that duality seems to imply a
deep connection between the neutrino $\bar{\nu}_{e}$ and the magnetic
monopole.

\bigskip\ 

\noindent \textbf{5.- Final Remarks}

\smallskip\ 

In this work we have shown that duality at the level of fundamental
constants leads to some interesting and intriguing conclusions: the Planck
time is a minimal time, the cosmological constant is quantized and the
magnetic monopole mass is related to the neutrino mass. One should expect
similar observations if the duality concept is applied to other physical
scenarios.

A question arises whether this duality of the fundamental constants might
shed some light on the controversy about the variability of fundamental
constants. Let us write formula (19) (for $n=1$) as

\begin{equation}
\frac{\mathcal{L}}{R}\frac{l}{l_{p}}=\frac{1}{2}.  \label{34}
\end{equation}%
We observe that this is a duality relation between two dimensionless
constants $\frac{\mathcal{L}}{R}$ and $\frac{l}{l_{p}}$. Similarly,
considering the ratios $\frac{m_{g}}{m_{e}}$ and $\frac{g^{2}}{e^{2}}$, one
sees that (25) is a duality expression between two dimensionless constants.
Of course, exactly the same conclusion can be obtained from the Dirac's
quantization condition (14), since in that case one may write (for $n=1$)

\begin{equation}
\frac{g^{2}}{\hslash c}\frac{e^{2}}{\hslash c}=\frac{1}{4}.  \label{35}
\end{equation}%
These observations mean that, from the point of view of duality symmetry
what seems to be essential are the dimensionless constants rather than the
dimensional ones, in agreement with Dirac's argument [3] and Duff's reply
[2]. In fact, it is easy to see that duality in terms of fundamental
dimensional constants does not make sense. For instance, let us assume a
duality for the light velocity $c$ of the form

\begin{equation}
c^{2}\leftrightarrow \frac{1}{c^{2}}.  \label{36}
\end{equation}%
If we set $c=1$ then this symmetry is lost. Thus, in order to maintain the
duality symmetry of an underlaying theory, it is necessary to express it in
terms of dimensionless constants. In turn, this implies that what matters is
the variability of such dimensionless constants, rather than dimensional
constants. Considering this observation, we discover that (34) and (35)
establish that time variability of a dimensionless fundamental constant
implies a time variability of its corresponding dual.

Now, one should expect that the duality of the dimensionless fundamental
constants is reestablished in a duality at the level of the fundamental
field theory. Maxwell field theory, with both electric and magnetic sources,
offers an excellent example of this remark. Therefore, one should be
interested in applying the ideas discussed in this paper in a corresponding
field theory in which duality may play a fundamental role. In fact, the
duality for linearized gravity used in Sec. 3 as starting point in
connection with the duality of the cosmological constant is a good example
of this idea. However, one may still be more ambitious and ask for a theory
in which duality acts as a fundamental principle. In a sense, this is the
principle suggested by the interconnection between the various string
theories leading to the so-called M-theory [14]. Thus, one may say that
M-theory is the final goal of a duality principle. The fine point is that
this idea may require a new and unexpected mathematical framework for its
realization. In a series of works [15]-[22], it has become more evident that
a candidate for such a mathematical framework is the oriented matroid theory
[23]. Hence, one of our aims for further research is to use the oriented
matroid theory as a mathematical tool in order to have a better
understanding of the duality of fundamental constants.

The main idea of the present work was to link duality symmetry with various
fundamental constants. In this respect, it is worth mentioning that a
relation between the cosmological constant and atomic units has been
established a long time ago [24]. In fact, this relation seems to present
some kind of duality between the cosmological constant similar to the
present discussion. Therefore, it may be interesting for further research to
analyze the ideas of Ref. 24 from the point of view of the present work.
Furthermore, there will be effects of duality symmetry in connection with
fundamental constants, and in particular with the cosmological constant,
which we might hope to be able to measure. In this sense the cosmic
geophysical observations discussed in Ref. 25 may be a guide, and this is
something we hope to consider in the near future.

From the present work the following natural questions may emerge:

(i) The expression (1) for the radiated power of gravitational waves is
calculated in linearized GRT, i.e., for weak gravitational fields. What
sense does it make then to bring it into context with the Planck time which
governs extremely strong gravity?

(ii) What does it mean to quantize a fundamental constant, as motivated by
some formal analogy for the cosmological constant? Wouldn't it be a proposal
against the spirit of such a constant?

(iii) Is there any physical meaning of the sum over all infinite energy
levels of the Hydrogen atom?

\noindent It is clear that, although these questions are interesting, their
answer might not be so simple. Nevertheless, it is tempting to try giving a
possible answer. Let us first discuss the question (i). It turns out that
exactly the same question can arise in the case of weak/strong coupling
duality of linearized gravity [4,26]. The answer in this case may rely on
the assumption of dual `phases' of M-theory: one which describes weak
gravity and the other, strong gravity. And each one would have their own
field theory limit. But the idea is that the M-theory itself becomes
invariant under a weak-strong duality transformation. From this perspective,
it seems surprising that one may touch this idea of dual phases of M-theory
by simply considering the duality between the maximum radiated power of
gravitational waves and Planck time. A similar argument can be applied in
the case of question (ii). M-theory should have two dual phases each one
with small/large cosmological constant. So, the traditional spirit of the
cosmological constant comes from just one of these dual phases, but as soon
as one realizes the possibility of the other dual gravitational phase then
the quantization of the cosmological constant becomes as a consequence. It
is worth mentioning that the idea of the quantum cosmological constant has
already appeared in other contexts [27,28]. At first sight it seems that the
question (iii) should correspond to a different scenario. However, since we
have assumed in Sec. 4 the weak/strong coupling duality for an Abelian gauge
theory, which is presumably part of M-theory, we find that a possible answer
might also be found in the concept of dual phases of M-theory. In fact,
suppose that we have a system in which in one phase can be described by the
associated constants $m_{e}$ and $e$ and in the other by $m_{g}$ and $g,$
respectively. In order for this description to make sense, something must
remain constant. According to formula (25) this is provided by the
combinations $m_{g}g^{2}$ and $m_{e}e^{2}$. Thus, such a constant must be
fundamental, dimensionless, and should not be related to any property of the
system itself. What other than the number $\pi $? It just happens that, as
the Weyler heuristic formula and the formula (27) indicates, such a constant
should be proportional to $\pi ^{2}$ rather than $\pi $ itself. Now, from
(27), one may obtain (29). The next step is simply to apply the famous
numerical series (30) due to Euler for determining the number $\pi ^{2}$.
What we obtain is the energy formula (31), which can be related to the
hydrogen atom. From this perspective, one has obtained the surprising result
that the quantum energy formula for the hydrogen atom is a consequence of
the dual phases of M-theory.

Although the above explanations in terms of the M-theory seem reasonable,
one can still have the feeling that the questions above require further
discussion. For instance, M-theory does not give an answer to the question:
What is the strong gravitational coupling phase? Attempts to answer this
question have been given by Nieto [4] and Hull [26]. In particular, Hull's
idea is to construct a theory from the dual gauge fields

\begin{equation}
D_{\mu \nu \alpha }=\epsilon _{\mu \nu \alpha \beta }h_{\alpha }^{\beta }
\label{37}
\end{equation}%
and 
\begin{equation}
C_{\mu \nu \alpha \gamma \rho \sigma }=\epsilon _{\mu \nu \alpha \beta
}\epsilon _{\gamma \rho \sigma \lambda }h^{\beta \lambda },  \label{38}
\end{equation}%
which are duals of the gravitational fluctuation $h$. Although these ideas
have generated some motivation (see Ref. 29 and references therein),
complete dual gravitational theory is still a mystery. Thus, since the
strong gravitational coupling phase is an open problem, one cannot expect to
give a general answer at the present to the above questions in terms of the
M-theory.

\bigskip\ 

\noindent \textbf{Acknowledgment: }This work was supported in part by the
UAS under the program PROFAPI-2006.

\smallskip\

\end{document}